\documentclass[preprint2]{aastex}

\usepackage{natbib}

\newcommand{\lsim}{\,\rlap{\raise 0.35ex\hbox{$<$}}{\lower 0.7ex\hbox{$\sim$}}\,}
\newcommand{\gsim}{\,\rlap{\raise 0.35ex\hbox{$>$}}{\lower 0.7ex\hbox{$\sim$}}\,}

\def \apjl{Astrophys.~J.}
\def \apjs{Astrophys.~J.~Supp.}
\def \apj{Astrophys.~J.}

\def \mnras{Mon.~Not.~Roy.~Astron.~Soc.}

\def \nat{Nature}

\def \grb{GRB~101219B\ }

\def \fermi{{\it Fermi\ }}
\def \swift{{\it Swift\ }}

\shorttitle{GRB\~101219B}
\shortauthors{Larsson et al.}

\begin{document}

\title{Evidence for jet launching close to the black hole in GRB~101219B - a \textrm{Fermi} GRB dominated by thermal emission}


\author{J.~Larsson\altaffilmark{1}, J.~L.~Racusin\altaffilmark{2} and J.~M.~Burgess\altaffilmark{1}}


\altaffiltext{1}{KTH, Department of Physics, and the Oskar Klein Centre, AlbaNova, SE-106 91 Stockholm, Sweden}
\altaffiltext{2}{NASA, Goddard Space Flight Center, Greenbelt, MD 20771, USA}

\begin{abstract}

We present observations by the {\it Fermi Gamma-Ray Space Telescope} Gamma-Ray Burst Monitor (GBM) of the nearby ($z=0.55$) GRB~101219B. This burst is a long GRB, with an associated supernova and with a blackbody (BB) component  detected in the early afterglow observed by the \swift X-ray Telescope (XRT).  Here we show that the prompt gamma-ray emission has a BB spectrum, making this the second such burst observed by \fermi GBM. The properties of the BB, together with the redshift and our estimate of the radiative efficiency, makes it possible to calculate the absolute values of the properties of the outflow. We obtain an initial Lorentz factor $\Gamma=138\pm 8$, a photospheric radius $r_{\rm{phot}}=4.4\pm 1.9 \times 10^{11}$~cm and a launch radius $r_0=2.7\pm 1.6 \times 10^{7}$~cm. The latter value is close to the black hole and suggests that the jet has a relatively unobstructed path through the star. There is no smooth connection between the BB components seen by GBM and XRT, ruling out the scenario that the late emission is due to high-latitude effects.  In the interpretation that the XRT BB is prompt emission due to late central engine activity, the jet either has to be very wide or have a clumpy structure where the emission originates from a small patch. Other explanations for this component, such as emission from a cocoon surrounding the jet, are also possible. 

\end{abstract}


\section{INTRODUCTION}
\label{intro}

Insight into the physical properties of jets in gamma-ray bursts (GRBs) is dependent on identifying the emission mechanism responsible for the prompt gamma-ray emission. This is an unsolved problem, where no theoretical model has yet managed to explain the full range of observed parameters for the smoothly broken power-law shape of the spectra. In particular, it is well known that synchrotron emission cannot explain the hard low-energy slopes of many bursts \citep{Crider1997,Preece1998}, while the majority of GRB spectra are significantly wider than expected for the simplest cases of thermal emission from the jet photosphere. From observations with \fermi there is growing evidence that emission from the jet photosphere is important in at least some bursts, although it is usually seen to co-exist with a dominant non-thermal component \citep{Ryde2010,Guiriec2011,Axelsson2012,Guiriec2013,Burgess2014a}.

Due to the difficulties in discriminating between different emission models, observations of GRB spectra which are close to pure Planck functions are very important. There have previously been a few GRBs with such spectra observed by BATSE \citep{Ryde2004} as well as one observed by \fermi GBM \citep{Ghirlanda2013}. For these bursts, however, the redshifts and radiative efficiencies are unknown, leading to uncertainties in the absolute values of outflow parameters derived from the spectral fits. Here we present observations of GRB~101219B, the first GRB with a spectrum that is well described by a Planck function that also has a detected afterglow and a known redshift. 

The prompt emission of \grb was observed by both \swift Burst Alert Telescope (BAT) and \fermi GBM \citep{Cummings2010,vanderHorst2010}, and the afterglow was detected from X-rays to optical energies by several observatories \citep{Gelbord2010,Olivares2010,deUgarte2011}. The optical signature of an accompanying Supernova (SN) was also detected \citep{deUgarte2011,Olivares2011} and the redshift was determined to be 0.55. An analysis of the SN spectrum by \cite{Sparre2011} showed it to be a broad-line type Ic, with properties similar to SN~1998bw, the archetypal GRB-SN \citep{Galama1998}. In addition, the \swift XRT spectra of the early afterglow revealed the presence of a highly significant BB component \citep{Starling2012}. Such a component has also been found in a small number of other GRBs and the origin has been attributed variously to the SN shock breakout, emission from the cocoon surrounding the jet, or late emission from the jet itself \citep{Starling2012,Sparre2012,Friis2013}. 

This paper is organized as follows: we present the spectral analysis of the gamma-ray observations in section \ref{gammaobs}, estimate the radiative efficiency from the \swift XRT observations in section \ref{effic} and derive the outflow parameters in section \ref{fireball}. We discuss our results in section \ref{discussion}. Throughout this paper we assume a flat Universe with $H_0=71\ \rm{km\ s^{-1}}$ and $\Omega_{\rm{M}} =0.27$. Error bars correspond to 1 sigma unless otherwise stated.

\section{OBSERVATIONS AND ANALYSIS} 
 
\subsection{Gamma-ray spectral analysis}
\label{gammaobs}

\begin{figure}
\begin{center}
\rotatebox{360}{\resizebox{90mm}{!}{\includegraphics{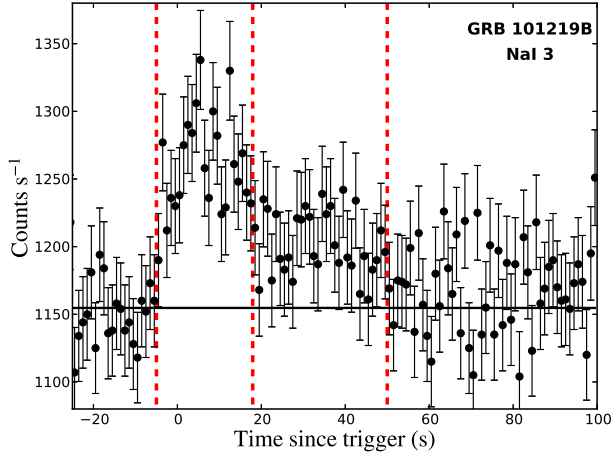}}}
\caption{Light curve of \grb\  from NaI~3 with 1~s bins. The two time intervals used for the spectral analysis are shown as dashed lines. The black line shows the background level, which was determined by fitting the light curve before $-10$~s and after $70$~s with a straight line.}
\label{lightcurve}
\end{center}
\end{figure}

In the analysis of the prompt phase of \grb we focus on the data from GBM \citep{Meegan2009}, which has a stronger signal as well as a wider energy range than the BAT.  We do check the BAT data for consistency though.  \grb is a relatively weak burst with a fluence of $3.99\pm 0.05 \times 10^{-6}\ \rm{erg\ cm^{-2}}$ in the $10-1000$~keV energy band and duration  $T_{90} =51.0 $~s (\citealt{vonkienlin2014}).  The light curve of the NaI detectors ($8-1000$~keV) is dominated by a single peak, as shown in Fig.~\ref{lightcurve}.

For the spectral analysis we use the NaI~3, 6 and 7 detectors, which all have angles with respect to the burst of \hbox{$<60^{\circ}$} . There is no signal in the BGO detectors (200 keV - 40 MeV) or in the Large Area Telescope (LAT, 30~MeV - 300~GeV).  The fits were performed using {\scriptsize XSPEC~v12.8.1} and pgstat statistics. To select time intervals for the spectral analysis we binned the light curve using the  Bayesian Blocks technique (as implemented in {\scriptsize astroML}, \citealt{vanderPlas2014}), which resulted in two time intervals, shown in Fig.~\ref{lightcurve}. 

\begin{figure}
\begin{center}
\rotatebox{270}{\resizebox{!}{80mm}{\includegraphics{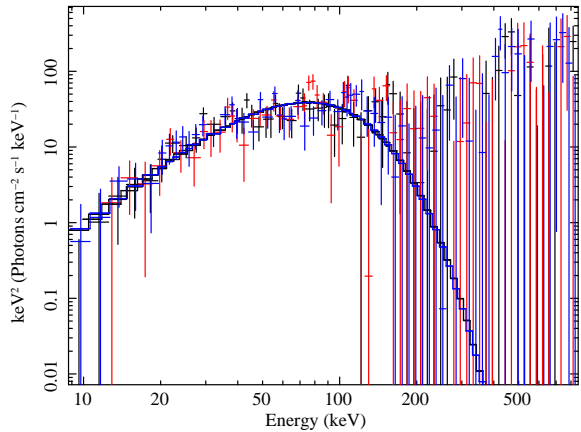}}}
\caption{Fermi GBM spectrum of \grb\ from the first interval in Fig.~\ref{lightcurve} fitted with a BB model. Data from Nai 3, 6 and 7 are shown in black, red an blue, respectively.}
\label{spectrum}
\end{center}
\end{figure}

GRBs are typically well described by the so-called Band function \citep{Band1993}, which is a smoothly broken power law characterized by a peak energy ($E_{\rm{peak}}$) and power-law indices $\alpha$ and $\beta$ below and above the peak, respectively. The results from fitting \grb with a Band function are listed in Table~\ref{fitparams}. The fits show that $\alpha$ is very hard, $0.63\pm 0.05$ and $0.88^{+0.24}_{-0.26}$ in the first and second interval, respectively, and close to the Rayleigh-Jeans value of 1. For comparison,  the median value of $\alpha$ for long GRBs in the GBM catalog is $-0.89$ \citep{Gruber2014} and synchrotron emission in the slow-cooling regime has $\alpha = -2/3$ (\citealt{Preece1998}, see also \citealt{Burgess2014b}). In addition, this burst also has an unusually low $E_{\rm{peak}}$ ($\sim 70$~keV in the first time interval), which is at the lowest 10th percentile of constrained $E_{\rm{peaks}}$ for GBM bursts  \citep{Gruber2014}, and even more exceptional when considering the low redshift. Motivated by these unusual spectral properties we attempt to fit the data with a BB model, which we find provides a good fit to the spectra. The fit to the first time interval is shown in Fig.~\ref{spectrum} and the best-fit parameters are presented in Table \ref{fitparams}. The temperature of the BB is seen to decrease between the two time intervals. 

In order to test the hypothesis that the spectrum of this burst, which has rather poor signal-to-noise, could still be consistent with slow-cooling synchrotron emission we use {\scriptsize XSPEC} to simulate $10,000$ spectra with $\alpha=-2/3$ and $\beta$ and $E_{\rm{peak}}$ taken from the fit to the first interval. The simulations are based on the background spectrum, response files and exposure time of the real data, and uses Poisson statistics to create synthetic spectra. Fitting these spectra with a Band function gives best-fit values of $\alpha > 0.63$ in only six cases, corresponding to a probability of $6\times 10^{-4}$ of obtaining such a hard value of $\alpha$ by chance. The hypothesis that $\alpha=-2/3$ can thus be rejected with very high confidence.

Even for the case when the prompt emission is completely dominated by the jet photosphere the observed spectrum is not expected to be a pure Planck function. Geometric effects are expected to broaden the spectrum, with the hardest $\alpha$ values of $+0.4- +0.5$ being obtained in the case of a wide jet viewed along the line of sight \citep{Peer2008,Beloborodov2011,Lundman2012,Deng2014}. In addition, we expect some broadening due to time-evolution of the spectrum during the $\sim 20-30$~s time intervals from which the spectra were extracted (it is not possible to obtain good constraints on spectral parameters on a shorter time scale for this burst). A comparison of the Band and BB fits can be used to constrain the amount of broadening in GRB~101219B. While $\alpha$ is close to the value expected for a pure BB, $\beta = -2.60$ is harder than the exponential cutoff of a BB. In line with this, a fit with a cutoff power-law model over-predicts the data at low energies as it needs to compensate for the somewhat harder high-energy slope. The flux derived from the BB fit to the first time interval is approximately $60 \%$ of the flux derived from the Band fit, which places an upper limit on the amount of broadening. As already mentioned, some of this will be accounted for by geometric effects and temporal evolution, but a small contribution from non-thermal emission, such as Comptonization of the BB, is also possible. 

For consistency we also analyzed the \swift BAT data of the burst. Spectra were created using standard BAT software (HEASOFT 6.15 and {\scriptsize BATGRBPRODUCT}). The results of the joint spectral fits are fully consistent with those presented above. We note, however, that the BAT spectra on their own are adequately fit with a simple power law ($\Gamma=-1.7\pm0.2$ in the first time interval). This is due to the combination of a weak signal and the limited band pass ($15-150$~keV), which means that the power law simply provides an average of the spectrum covering the spectral peak. 

\subsection{Radiative efficiency and jet opening angle}
\label{effic}

\begin{figure}
\begin{center}
\rotatebox{90}{\resizebox{!}{80mm}{\includegraphics{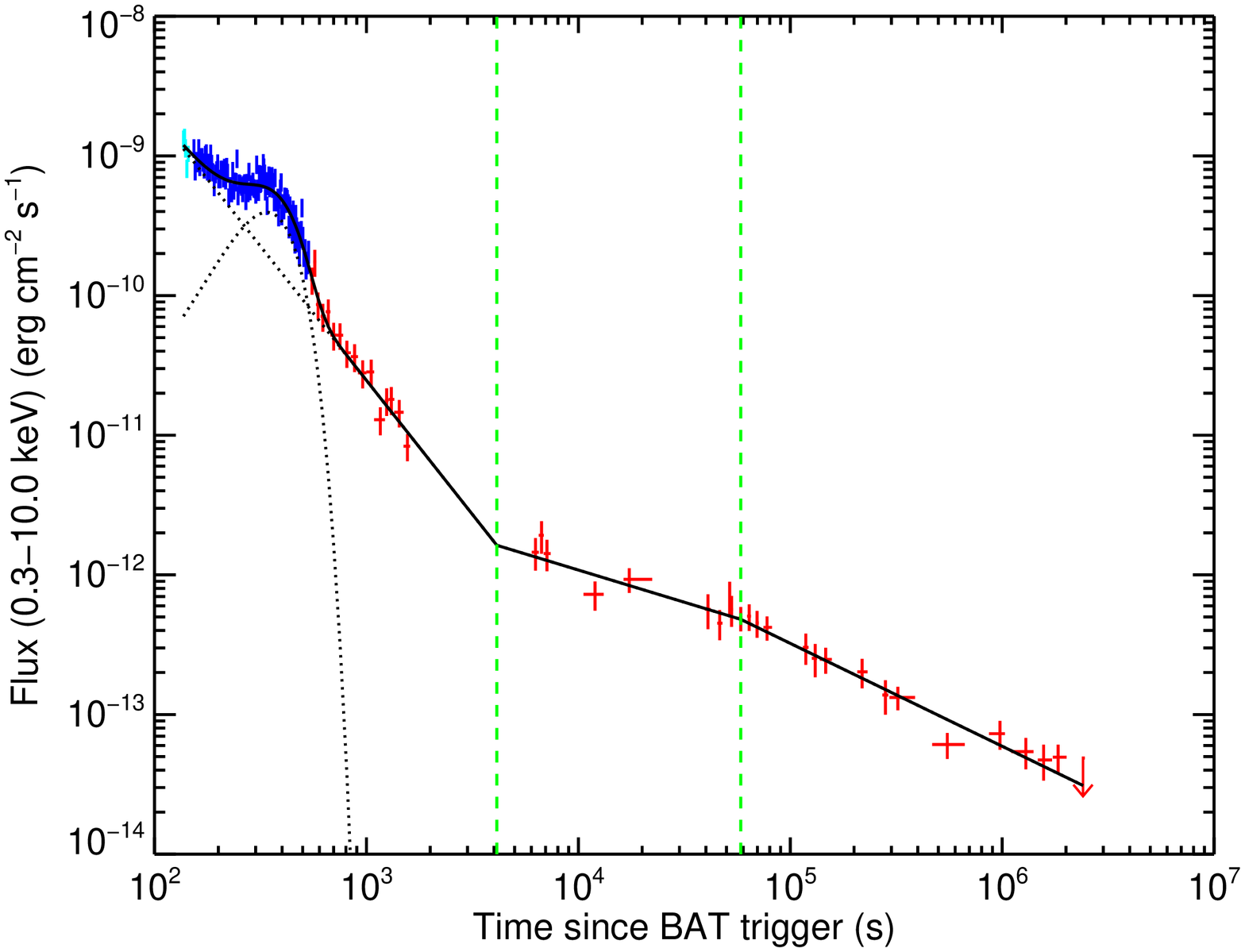}}}
\caption{\swift XRT light curve of GRB~101219B. Cyan data points were taken in WT settling mode, blue points in WT mode and red points in PC mode. The solid black lines shows the fit with three power-law segments, separated by breaks marked by green dashed lines. The dotted black line shows a flare that was removed from the fit.}
\label{xrtlc}
\end{center}
\end{figure}

The radiative efficiency is defined as $\eta = E_{\gamma, \rm{iso}} / (E_{\gamma, \rm{iso}} + E_{\rm{k}})$, where $E_{\gamma, \rm{iso}}$ is the isotropic energy emitted in gamma rays and  $E_{\rm{k}}$ is the kinetic energy of the jet. A measurement of $\eta$ is needed in order to calculate the absolute values of fireball parameters from the BB properties (see section \ref{fireball}). The value of $E_{\rm{k}}$  can be derived from X-ray afterglow observations, as described by \cite{Zhang2007}.  We obtained the X-ray afterglow light curve and spectra for \grb from the \swift XRT repository \citep{Evans2009}. Fitting the count rate light curve using the method of \cite{Racusin2009} yields three power-law segments with temporal decay indices of $\alpha_1=1.92\pm0.07$, $\alpha_2=0.46^{+0.13}_{-0.22}$, $\alpha_3=0.74^{+0.13}_{-0.09}$ with two breaks at $t_{\rm{br,1}}=4.1^{+1.3}_{-0.9}$~ks and $t_{\rm{br,1}}=58^{+57}_{-36}$~ks. These segments can be identified as late prompt emission, a shallow phase with continuous energy injection and the normal decay due to the decelerating fireball. No jet break is detected. A flare around 300-600~s was modeled by a Gaussian. The light curve and the fit are shown in Fig.~\ref{xrtlc}.

We estimate $E_{\rm{k}}$ from the normal decay phase, assuming typical values for the fraction of energy in electrons and magnetic fields ($\epsilon_e=0.1$ and $\epsilon_b=0.01$) and neglecting Inverse Compton emission. We first determined the intrinsic column density, $N_{\rm{H,int}}$, by fitting the full spectrum during the normal decay (i.e.~after 58~ks) with a power law modified by intrinsic and Galactic absorption (with the Galactic absorption fixed at $3.1\times 10^{20}\ \rm{cm}^{-2}$). This yields  $N_{\rm{H,int}}=1.2^{+1.2}_{-1.0}\times 10^{21}\ \rm{cm}^{-2}$, consistent with the results of \cite{Starling2012}. We then fit the model with $N_{\rm{H,int}}$ fixed at this value to the spectrum at 58~ks and use the fit results together with equation (5) in \cite{Racusin2011} to derive $E_{\rm{k}}$. Assuming slow-cooling synchrotron emission with an electron spectral index $p>2$, the observed spectrum is consistent with being above the synchrotron cooling frequency. We obtain a value of  $E_{\rm{k}} = 6.4 \pm 3.5 \times 10^{52}$~ergs. With $E_{\rm{\gamma, iso}} = 3.4\pm0.2 \times 10^{51}$~ergs (over $10-1000$~keV in the rest frame of the burst) this corresponds to a relatively low efficiency of $\eta =5\pm 2 \% $.

This value of $\eta$ is associated with systematic uncertainties from our assumptions, in particular regarding the values of  $\epsilon_e$ and $\epsilon_b$. Although the values of $\epsilon_b$ reported in the literature span many orders of magnitude, we note that our estimate of $E_{\rm{k}}$ is only weakly dependent on this parameter. As an example, assuming a much lower value of  $\epsilon_b = 5.5 \times 10^{-6}$ (the median value found by \citealt{Santana2014} for $p=2.8$ and a constant density medium) gives  $\eta \sim 2 \%$, i.e.~only slightly lower than our estimate above. The dependence on $\epsilon_e$ is stronger, but measurements of this parameter have a rather narrow distribution, with $\sim 60\%$ of GRBs falling in the range $0.1-0.3$ \citep{Santana2014}, in line with theoretical simulations \citep{Sironi2011}. Increasing our value of $\epsilon_e$ to 0.3 at the high end of this distribution gives $\eta \approx 20 \%$, which gives an estimate of the uncertainty from this parameter. 

A limit on a jet break can be obtained from the last time when such a break could be present without being detectable (see \citealt{Racusin2009}). For \grb this time is $t_{{\rm{b}}}>15.3$~days. Assuming an ambient number density of $1~\rm{cm^{-3}}$ and $\eta$ and $E_{\gamma, \rm{iso}}$ from above, this corresponds to a limit on the jet opening angle of $\theta_{\rm{jet}}>17.1^{\circ}$ and a collimation corrected energy  $E_{\gamma} > 1.5\times 10^{50}$~ergs.

\subsection{Fireball parameters}
\label{fireball}

The observed temperature and flux of the BB, together with $\eta$ and $z$, can be used to calculate the properties of the outflow in the standard fireball scenario. Following \cite{Peer2007} we calculate the Lorentz factor ($\Gamma$), the radius where the fireball starts accelerating ($r_{0}$), the saturation radius ($r_{\rm{s}}$, where the acceleration stops)  and  the radius of the photosphere ($r_{\rm{ph}}$). In the calculations we assume that there is no dissipation below the photosphere and that $\Gamma$ is much larger than the inverse of the opening angle of the jet. The results for the two intervals are presented in Table~\ref{partable}. 

The main source of uncertainty in these results is the radiative efficiency, as already discussed above. Another uncertainty is the contribution of non-thermal emission to the gamma-ray spectrum. A very conservative limit on this can be obtained by comparing the fluxes from the Band and BB fits, as described in section \ref{gammaobs}. Taking the difference of these fluxes to be non-thermal emission would reduce the values of  $r_{0}$ and $r_{\rm{s}}$ by about a factor of two, while at the same time increasing the values of $r_{\rm{ph}}$ and $\Gamma$ by about $15\%$, as compared to the values presented in Table~\ref{fitparams}. 

 \begin{deluxetable}{ccccccccc}
\tabletypesize{\scriptsize}
\tablecaption{Spectral fits and Fireball parameters \label{partable}}
\tablewidth{0pt}
\tablehead{
\colhead{Time$^{\rm{a}}$} & \colhead{Band $\alpha^{\rm{b}}$} & \colhead{Band $\beta^{\rm{b}}$} & \colhead{Band $E_{\rm{peak}}^{\rm{b}}$} & \colhead{BB kT$^{\rm{c}}$} & \colhead{$r_{0}^{\rm{d}}$} & \colhead{$r_{\rm{s}}^{\rm{d}}$}  & \colhead{$r^{\rm{d}}_{\rm{ph}}$ } & \colhead{$\Gamma^{\rm{d}}$}\\
\colhead{(s)} & & & \colhead{(keV)} & \colhead{(keV)} &  \colhead{(cm)} &  \colhead{(cm)}  & \colhead{( cm)}  & }
\startdata
-5--18 & $0.63\pm 0.05$ & $-2.60^{+0.18}_{-0.37}$  & $72.9\pm 14.8$  &$19.1^{+0.7}_{-0.6}$ & $2.7\pm 1.6 \times 10^{7}$ & $3.7\pm 2.3 \times 10^{9}$ &  $4.4\pm 1.9 \times 10^{11}$ &  $138\pm 8$ \\
18-50 & $0.88^{+0.24}_{-0.26}$ & $-2.23^{+0.12}_{-0.22}$ & $38.6^{+6.2}_{u} $  &$11.2^{+0.8}_{-0.7}$ &  $4.5\pm 2.8 \times 10^{7}$ & $4.1\pm 2.6 \times 10^{9}$ & $4.9\pm 2.3 \times 10^{11}$ &  $92\pm 8$ \\
\enddata
\tablecomments{}
\label{fitparams}
\tablenotetext{a} {Time interval relative to the GBM trigger.}
\tablenotetext{b} {Best-fit parameters from fitting the GBM data with a Band model.}
\tablenotetext{c} {Best-fit parameters from fitting the GBM data with a BB model.}
\tablenotetext{d} {Fireball parameters derived from the BB fits.}
\tablenotetext{u}{Parameter error range unconstrained.}
\end{deluxetable}


\section{DISCUSSION}
\label{discussion}

As noted in Section~\ref{intro} there is growing evidence for photospheric emission in the prompt phase of GRBs, but very few examples where the spectra are close to pure BBs. It is clear that special conditions are required in order to obtain such spectra. First, a fairly wide jet viewed close to the line of sight is needed in order to avoid significant geometric broadening. This is in line with the limit on the jet opening angle from section \ref{effic}. However, it is interesting to note that the low-energy slope of \grb is  harder than predicted for a spherically symmetric wind (see section \ref{gammaobs}). A possible explanation for this is that the emission arises from a small patch within a clumpy jet, invalidating the assumption of a steady flow that is spherically symmetric inside $1/\Gamma$. An additional requirement for having a narrow spectrum is the lack of significant dissipation of energy in the jet.  

The outflow parameters in Table~\ref{fitparams} provide important information about the jet. It is particularly interesting to relate the base of the jet to the size of the presumed compact remnant. In the first time interval $r_{0} =2.7\pm 1.6 \times 10^7$~cm, which corresponds to ten times the event horizon for a  $18 \pm 11\ M_{\sun}$ maximally spinning  black hole (or a $9 \pm 5\ M_{\sun}$ non-spinning one). This mass range is close to the expected values for the generally assumed massive progenitor star, and suggests that the jet is launched from close to the black hole and then has an unobstructed path through the star. The latter requires a highly asymmetric explosion, in line with the observationally and theoretically well-established view that SN explosions are asymmetric. The only other GRB for which absolute values of the outflow parameters have been calculated is the highly energetic GRB~090902B,  where \cite{Peer2012} found $r_{0} = 3.0 - 7.4 \times 10^8$~cm, i.e. significantly larger than for GRB~101219B. That the conditions in this burst are different from \grb is also clear from the fact that the spectral shape is broader, most likely indicating subphotospheric dissipation.

\subsection{Connection with the blackbody component in the early afterglow}
\label{swiftbb}

\grb belongs to a small group of bursts in which a highly significant BB component has been detected in \swift XRT data of the early afterglow \citep{Starling2012}. Specifically, \cite{Starling2012} fit four time intervals between $180$ and $1080$~s after the GBM trigger (around the time of the flare in the light curve, see Fig.~\ref{xrtlc}) using a model comprising a power law and a BB, modified by absorption (see their table 5 for the full fit results). The BB has an initial rest-frame temperature of $0.3$~keV and then decreases with time.

\begin{figure*}
\begin{center}
\resizebox{80mm}{!}{\includegraphics{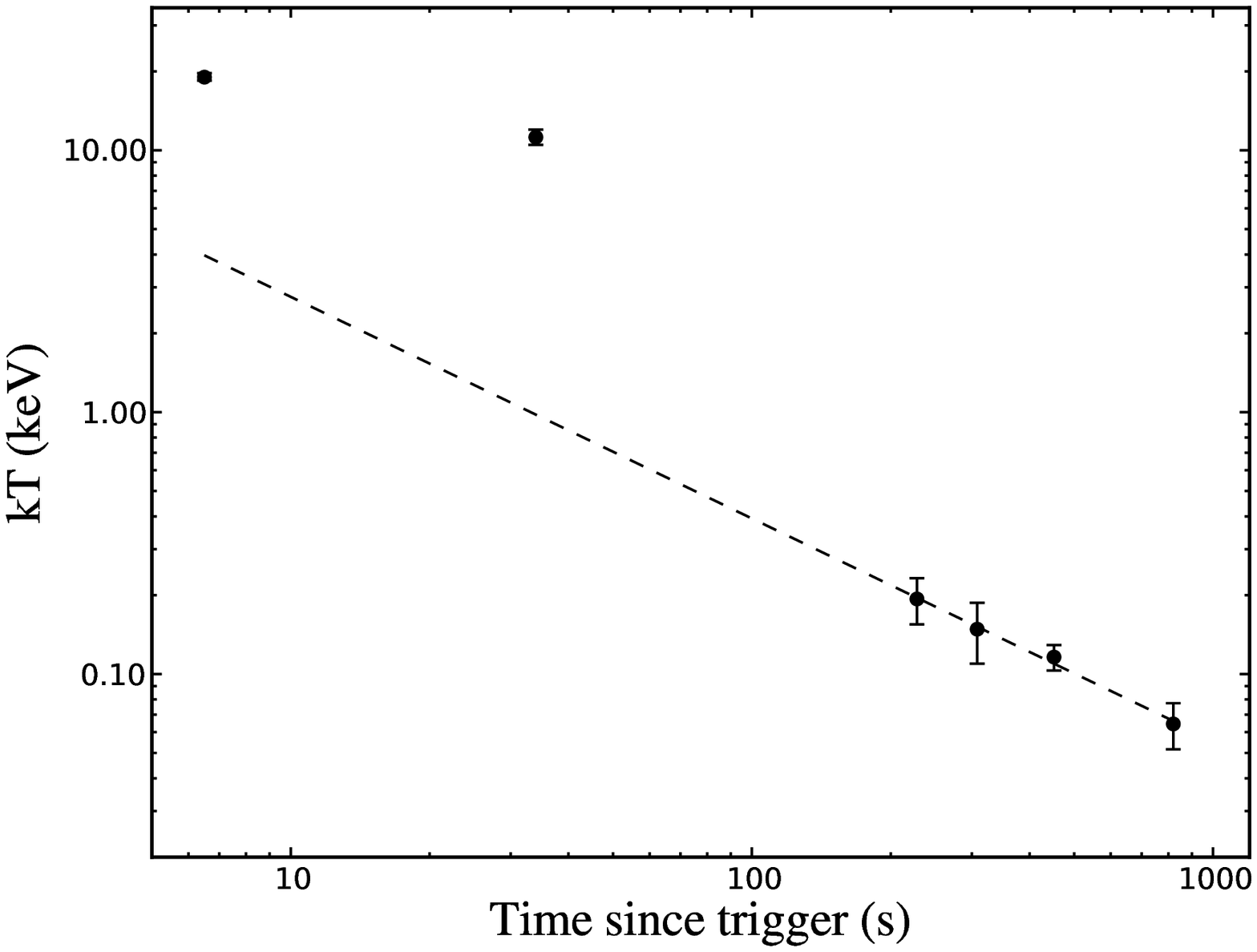}}
\resizebox{80mm}{!}{\includegraphics{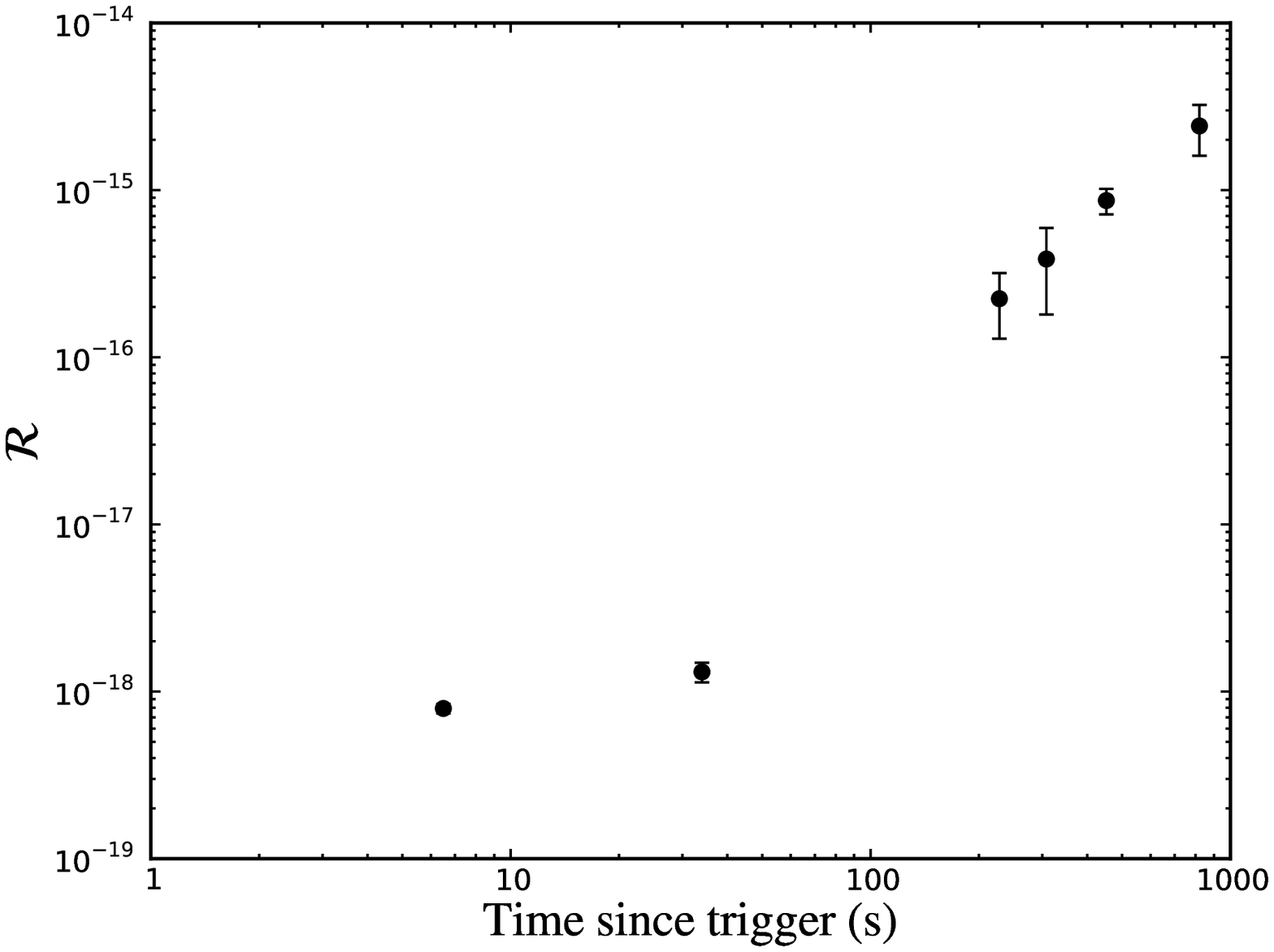}}
\resizebox{80mm}{!}{\includegraphics{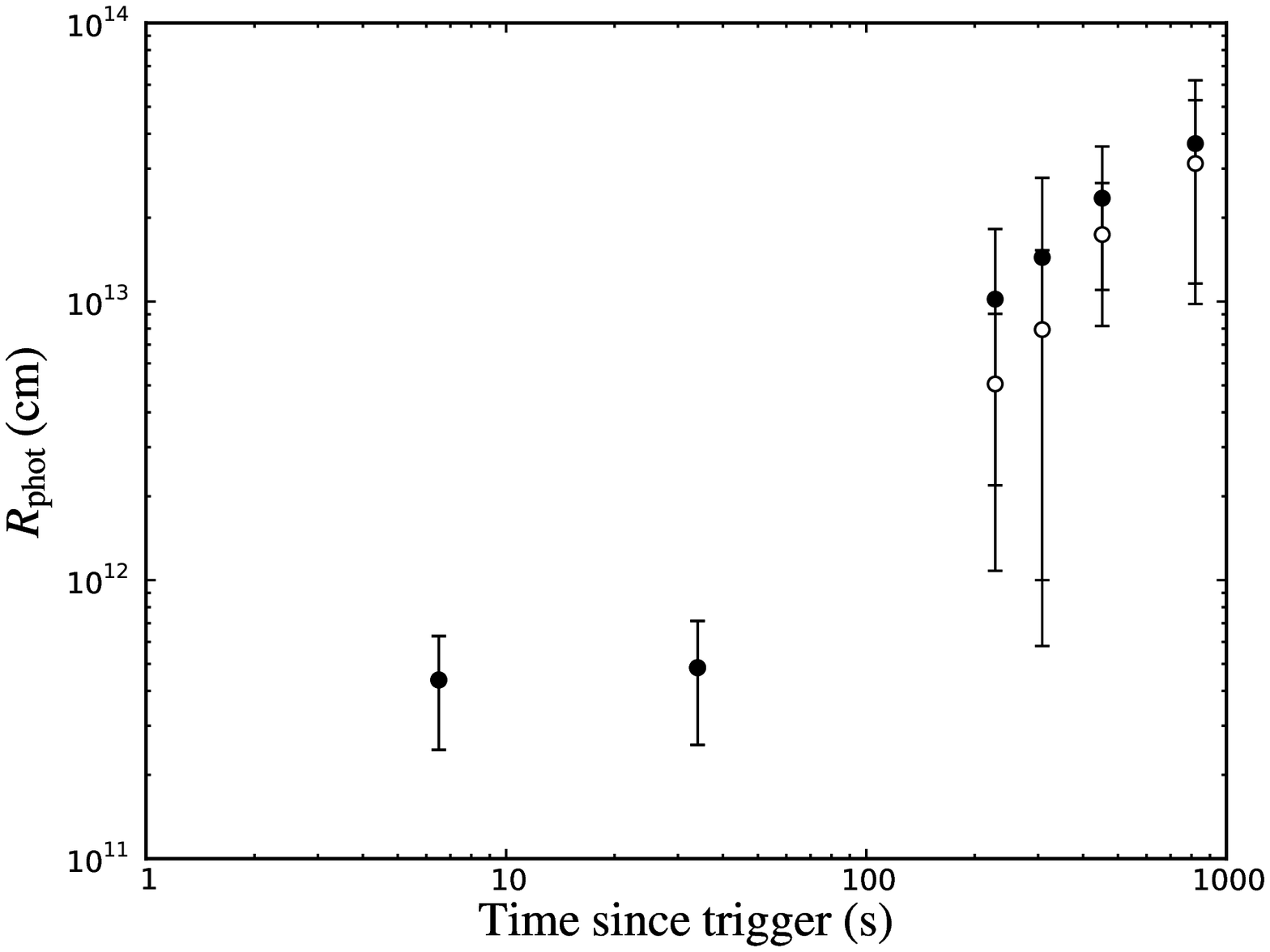}}
\resizebox{80mm}{!}{\includegraphics{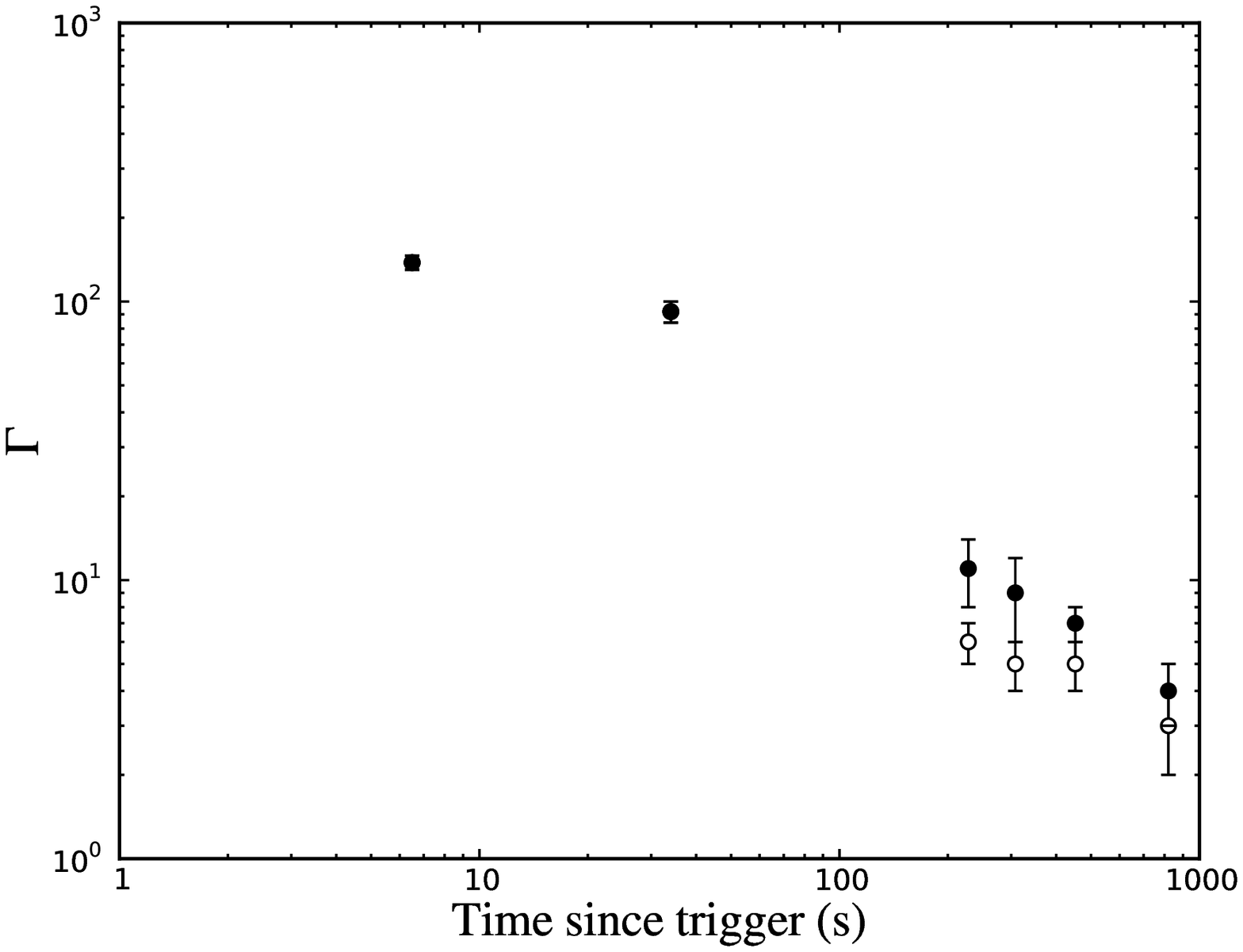}}
\caption{BB properties and outflow parameters obtained from fits to the prompt emission observed by \fermi GBM (first two data points) and the early afterglow observed by {\it Swift} XRT (last four points, from \citealt{Starling2012}). Times refer to the GBM trigger time. Upper left: observed BB $kT$. The dashed line is the fit to the XRT data points. Upper right: effective size of the photosphere $\mathcal{R} = (F_{\rm{BB}}/\sigma T^4)^{1/2}$. Lower, left: radius of the photosphere ($r_{\rm{ph}}$). Lower, right: Lorentz factor of the jet ($\Gamma$). The error bars on the XRT data points for  $\mathcal{R}$,  $\Gamma$ and $r_{\rm{ph}}$ assumes 10\% errors on the fluxes reported by \cite{Starling2012}. The solid points in the bottom panels correspond to the scenario that all the emission observed by \swift XRT is prompt emission from the jet, while the open symbols are for the assumption that only the BB component is prompt emission. }
\label{ktpars}
\end{center}
\end{figure*}

In order to investigate any possible connection between the BB in the prompt emission and the BB in the early afterglow we plot in Fig.~\ref{ktpars} the evolution of the BB temperature as a function of time for the two datasets.  The temporal evolution of the temperature in the \swift XRT data is well described by a power law (shown by the dashed line, with decay index $-0.85$), but the extrapolation of this fit to early times is clearly inconsistent with the GBM data. A broken power-law model also fails to connect the two data sets. In Fig.~\ref{ktpars} we also plot the parameter $\mathcal{R} = (F_{\rm{BB}}/\sigma T^4)^{1/2}$, which is related to the effective transverse size of the photosphere. This parameter is also seen to evolve differently during the two phases. 

The lack of a smooth connection between the observed properties of the BB in the two data sets, together with the fact that the BB flux actually increases at late times \citep{Starling2012},  rules out the scenario that the late emission is due to high latitude emission from the jet after the central engine has died. We therefore explore the scenario that the late emission is due to late central engine activity, and calculate the outflow parameters from the low-energy BB as described in section \ref{fireball}. The resulting values of $\Gamma$  and $r_{\rm{rphot}}$ are shown in Fig.~\ref{ktpars} for two scenarios: (i) all the late emission is prompt emission and (ii) the BB is the prompt emission from the jet, while the power-law component arises in a different region where the jet has started interacting with the circumstellar medium. The two scenarios produce comparable results. We see that at late times the jet has a low $\Gamma$ in the range $3-6$.  In order for such a slow jet to still produce a narrow BB spectrum the jet has to be very wide ($\gg 20^{\circ}$, consistent with the limit of           $\theta_{\rm{j}} > 17.1^{\circ}$ from section \ref{effic}) or, alternatively, the emission could originate from a small patch within a clumpy jet. 

As already discussed by \cite{Starling2012}, there are also other possibilities for the origin of the low-energy BB. In particular, they find that emission from a cocoon surrounding the jet can explain the main features of the emission, while shock breakout from a SN is disfavored from the large inferred radius ($\sim 10^{12} - 10^{13}$~cm, much larger than the $10^{11}$~cm expected from the presumed Wolf-Rayet progenitor). While breakout through a thick wind could accommodate the large radius in the SN scenario, the spectrum is in this case expected to be flat in $\nu F_{\nu}$ \citep{Svirski2014}, contrary to what is observed for GRB~101219B.

\acknowledgments

This work was supported the Swedish National Space Board.

\clearpage

\end{document}